\documentstyle[aps,preprint]{revtex}
\begin{document}
\draft

\title{First-Passage Time Distribution and Non-Markovian Diffusion Dynamics
of Protein Folding }
\author{${}^a$Chi-Lun Lee, ${}^b$George Stell, ${}^{bcd}$Jin Wang}
\address{${}^a$Department of Physics, State University of New York at Stony
Brook, Stony Brook, NY 11794 \\
${}^b$Department of Chemistry,
State University of New York at Stony Brook,
Stony Brook, NY 11794 \\
${}^c$Global Strategic Analytics Unit, Citigroup, One Huntington
Quadrangle, Suite 1N16, Melville, NY 11747 \\
${}^d$Department of Physics, Jilin University, Changchun, Jilin
130021, People's Republic of China}

\maketitle


\begin{abstract}
\tighten
We study the kinetics of protein folding via statistical
energy landscape theory. We concentrate on the local-connectivity
case, where the configurational changes can only occur among
neighboring states, with the folding progress described in terms
of an order parameter given by the fraction of native
conformations. The non-Markovian diffusion dynamics is analyzed in
detail and an expression for the mean first-passage time (MFPT)
from non-native unfolded states to native folded state is
obtained. It was found that the MFPT has a V-shaped
dependence on the temperature. We also find that the MFPT is
shortened as one increases the gap between the energy of the
native and average non-native folded states relative to the
fluctuations of the energy landscape. The second- and higher-order
moments are studied to infer the first-passage time (FPT)
distribution. At high temperature, the distribution becomes close
to a Poisson distribution, while at low temperatures the
distribution becomes a L\'evy-like distribution with power-law
 tails, indicating a non-self-averaging intermittent behavior
of folding dynamics. We note the likely relevance of this result
to single-molecule dynamics experiments, where a power law
(L\'evy) distribution of the relaxation time of the underlined
protein energy landscape is observed.
\end{abstract}

\newpage

\section{Introduction}
The study of diffusion along a statistical energy landscape is a
very important issue for many fields. In the field of protein
folding, the crucial question is how the many possible
configurations of polypeptide chain dynamically converge to one
particular folded state\cite{Levinthal}. Clearly, a statistical
description is needed for a large number of configurational
states. Recently, a new view of protein folding based on
energy-landscape theory was developed\cite{BW,BSOW,Shak,Dill,WOW}.
In this theory, there exists a global bias of the energy landscape
towards the folded state due to natural evolution selection.
Superimposed on this is the fluctuation or the roughness of the
energy landscape coming from conflicting interactions of the amino
acid residues. The resulting protein-folding energy landscape is
like a funnel. The funnel picture of folding is currently in
agreement with experiments\cite{Exp} and consistent with both
lattice and off-lattice simulations\cite{SO,Brooks,Shak1,WSW}.

It is very important to discuss the dynamics of folding and the
nature of the pathways on the funnel-like energy landscape.
Initially, there are multiple routes towards native folded states.
Due to the competition between roughness of the landscape and
entropy, down in the funnel, there may exist local glassy traps
(minima). Then discrete pathways leading to folding emerge. It is
crucial to determine the influence of the global bias towards
folded state on the actual folding process itself. Following the
study of the thermodynamics of the folding-energy landscape, the
kinetics of folding along the order parameter that represents the
progress of folding towards the native state can be discussed.
Although in the multi-dimensional state space, the states are all
locally connected, the singled-out order parameter (for example,
$\rho$, the fraction of native configurations ,or q, the fraction
of native contacts) which represents how close structurally the
protein is towards its native state, may or may not have local
connectivity. When the kinetic process is fast, either because of
the large thermodynamic driving force or because the process is
activationless, the native state can either be reached in one shot
or through some intermediates that are formed very rapidly ( the
unravelling from those intermediate states or traps is often
needed to reach the final native state). This is the case in which
the states in order-parameter space can move globally from one to
another in an essentially discontinuous way\cite{WSW}. On the
other hand, if the kinetics are slow due to the nature of the
activation folding process, then in general, the states are
locally connected in order-parameter space. The dynamic process
can be studied by a kinetic master equation, and in the local
connectivity limit, the kinetic equation reduces to a diffusion
equation.

When the energy landscape is smooth, the average diffusion time is
a good parameter for characterizing the dynamical process. On the
other hand, when the energy landscape is rough, there exist large
fluctuations of the energies, and the diffusion time is expected
to fluctuate very much around its mean. In that case the average
diffusion time is no longer a good parameter to characterize the
dynamics. One needs to know the full distribution of the diffusion
times in characterizing the folding process.

It is worth mentioning that the aforementioned problem has many
potentially important applications in other fields too. For
example, in considering glasses, viscous liquids, and spin
glasses\cite{spinglass}, the distribution of barriers, and
therefore the distribution of diffusion times, is crucial in
understanding the kinetics near the trapping or glass transition
temperature, where the energy landscape becomes rough.

In the field of protein dynamics, pioneering work by Frauenfelder
et al.\cite{frau} has shown experimentally that the energy
landscape is complex through the study of CO rebinding to Mb after
photo dissociation. Their work reveals evidence for conformational
substates that are organized in a hierarchical fashion. The study
of dynamics along this energy landscape is very important in
understanding the rebinding kinetics as well as the underlying
structure of the landscape. This would help to find the
interrelationships between structures, function, and dynamics of
proteins\cite{PNWW}. At low temperatures, the energy landscape
becomes rough, with the fluctuation of the landscape causing
non-self-averaging behavior in diffusion time. Here an
understanding of the distribution of the barriers, and therefore
the diffusion-time distribution, is necessary in order to
adequately characterize the kinetics of the whole system.

The average diffusion time along the folding funnel has been
studied both analytically\cite{BW,WSW,Shak2,LLSW} and
numerically\cite{Shak,Dill,SO}. In this paper, we obtain results
for the MFPT (mean first-passage time) as a function of the
temperature, the energy gap $\delta \epsilon$, and roughness
$\Delta \epsilon$ of the folding landscape by solving a
random-energy based model\cite{BW}. The MFPT is found to be
shorter when the ratio $\delta\epsilon/\Delta\epsilon$ is
increased. We investigate the fluctuation or variance and the
higher-order moments of FPT (first-passage time) as well as the
FPT distribution function. Above a kinetic transition temperature
$T_0$, the FPT is well behaved, and the distribution tends to be
Poissonian. On the other hand, as the temperature decreases, the
distribution of FPT starts to become broader around and below
$T_0$. FPT distribution develops a power-law tail, approaching a
L\'evy distribution, extending over large scale of time. The
non-self-averaging behavior of the kinetics gradually dominates.
Within this temperature regime the behavior of the distribution
function is often needed to adequately capture the whole kinetics.
By solving a corresponding unfolding process we also obtain a
folding transition temperature $T_f$, defined by the point where
the folding and unfolding curves intersect. The results show
strong correlations between $T_f$ and $T_0$.

It is worth pointing out that the reactions and activated barrier
crossings are stochastic events. The laws of chemical kinetics are
merely statistical laws describing the average behavior of
populations. It is now possible to measure the reaction dynamics
of individual molecules in the laboratory\cite{Moerner,Xie1}. This
opens the way for the statistics of reaction events to be directly
tested. When the traditional phenomenology based on simple rate
laws is valid for large populations, experiments on individual
molecules or small numbers of them should give simple statistics.
Generally, on complex energy landscapes such as biomolecules, the
populations often do not obey simple exponential decay laws, and
the activation processes often do not follow the simple Arrhenius
law. The study of the statistics of individual molecular reaction
events can greatly clarify these more subtle reaction processes.

At the level of large populations, the barrier-crossing picture
has been shown to often provide an adequate characterization of
the non-exponential kinetics usually seen in complex systems. The
dynamics of the barrier crossing on a fluctuating energy landscape
itself clearly leads to non-Poissonian statistics for individual
molecules. The environmental fluctuations lead to
``intermittency''\cite{WOW,OWW}. The intermittency reflects the
fact that certain relatively rare configurations of the
environments are the ones that most favor the reaction. This means
that the distribution develops fatty tails. The dynamics and
folding of proteins of single molecules also exhibits similar
behavior\cite{Chu}. In particular, the recent experiments carried
out on single-molecule enzymatic dynamics\cite{Xie} showed this
intermittency phenomena and the distribution of reaction time of
underline energy landscape of proteins approached L\'evy-like
distribution (or power-law decay).

We organize the paper as follows: We first establish the
theoretical foundations and give detail steps in studying folding
diffusion dynamics in the Theory section. Then in the section of
Results and Discussions we give a thorough discussions on the
results of both MFPT and distributions of FPT. The connection
between our theoretical results yielding a L\'evy-like
distribution of FPT and protein-dynamics experiments is also
discussed. At the end, we give the conclusions. An appendix is
added to give further details of the derivation of the diffusion
equation.

\section{Theory}
To describe the folding process, it is often convenient to define
an order parameter that characterizes the degree of folding. In
this paper, we use the fraction of native conformations or native
bond angles as the order parameter $\rho$. Furthermore, we assume
the local connectivity condition here, assuming that the dynamics
changes continuously with $\rho$.

The model we study here was introduced earlier\cite{BW,LLSW}. The
problem of protein folding dynamics can be illustrated as random
walks on a rough energy landscape. In this model, the energy
landscape is generated by a random energy model\cite{Derrida},
which assumes that the energies of non-native states and their
interactions are random variables with given probability
distributions. In this model polypeptide chain, there are $N$
residues, and for each residue there are $\nu+1$ allowed
conformational states. A simplified version of the Hamiltonian or
the protein energy function is:
\begin{equation}
  H = -\sum\epsilon_i(\alpha_i)-\sum J_{i,i+1}(\alpha_i,\alpha_{i+1})
  -\sum K_{i,j}(\alpha_i,\alpha_j)
\end{equation}
where the summation indices $i$ and $j$ are labels for the
residues in a polypeptide chain, and $\alpha_i$ represents the
conformational state of the $i$th residue. The first term
represents the one-body potential. The second term represents the
interactions of nearest neighbors in sequence, and the last term
represents the two-body interactions of amino acids distant in
sequence but close in space. In this random energy model, the
energy terms $\epsilon_i, J_{i,i+1},$ and $K_{i,j}$ for native
states are fixed to be $\epsilon_0, J_0,$ and $K_0$, respectively,
whereas for non-native states they are generated by independent
random variables. For simplicity the probability distributions of
these random variables are assumed to be Gaussians with means
$\bar{\epsilon}$, $\bar{J}$, $\bar{K}$ and widths
$\Delta\epsilon$, $\Delta J$, $\Delta K$ separately. Therefore the
probability distribution function $P(E, N_0)=<\delta(E-H)>$ for a
protein with total energy $E$ and $N_0$ native amino acids
(therefore $\rho=N_0/N$) is also a Gaussian with mean
\begin{eqnarray}
  \bar E(N_0) &=& -N\left[\frac{N_0}{N}\epsilon_0 + \left(\frac{N_0}{N}\right)^2
  L_0 + \left(1-\frac{N_0}{N}\right)\bar\epsilon + \left(1-\left(\frac{N_0}{N}
  \right)^2\right)\bar L\right] \nonumber \\
  &=& -N\left[\rho\epsilon_0 + \rho^2 L_0
  + (1-\rho)\bar\epsilon + (1-\rho^2)\bar L\right]
\end{eqnarray}
and width
\begin{eqnarray}
  \Delta E(N_0) &=& \left\{N\left[\left(1-\frac{N_0}{N}\right)\Delta \epsilon^2
  + \left( 1-\left(\frac{N_0}{N}\right)^2\right) \Delta
  L^2\right]\right\}^{1/2}
  \nonumber \\
  &=& \left\{N\left[(1-\rho)\Delta \epsilon^2
  + ( 1-\rho^2) \Delta L^2\right]\right\}^{1/2},
\end{eqnarray}
where $L_0 \equiv J_0 + zK_0$, $\bar L \equiv \bar J + z\bar K$,
and $\Delta L^2 \equiv \Delta J^2 + z\Delta K^2$. $z$ is the
average number of neighbors surrounding a residue distant in
sequence. Since the spatial collapse is usually fast compared with
the rest of the folding process, this dynamical effect is ignored
here for simplicity and it is assumed here that $z$ is a constant
throughout the folding process. By this random-energy construction
one can easily generate energy surfaces with roughness controlled
by $\Delta \epsilon$ and $\Delta L$ and global bias determined by
$\delta \epsilon \equiv \bar\epsilon - \epsilon_0$ and $\delta L
\equiv \bar L - L_0$. Further simplification is made by the
assumption that different protein conformational states are
uncorrelated. This independence assumption also provides a way of
introducing large fluctuations onto the energy landscape. It is
also likely that by applying this assumption one has already
brought in some ingredients of cooperativity. With this
approximation the average free energy and other thermodynamic
quantities can be derived through the use of the microcanonical
ensemble analysis. The result is $F(\rho) = E' - T S$, where $T$
is a scaled temperature,
\begin{equation}
  E' = -N \left[(\bar\epsilon+\bar L) + \frac{\Delta\epsilon^2 + \Delta L^2}{T}
  + \left(\delta\epsilon-\frac{\Delta\epsilon^2}{T}\right)\rho
  + \left(\delta L-\frac{\Delta L^2}{T}\right)\rho^2\right],
\end{equation}
is the energy function of the system and
\begin{equation}
  S = -N \left[ \rho\log\rho + (1-\rho)\log\left(\frac{1-\rho}{\nu}\right)
  + \frac{\Delta\epsilon^2(1-\rho)+\Delta L^2(1-\rho^2)}{2T^2}\right]
\end{equation}
is the entropy of the system.

The kinetic folding process is approximated by the Metropolis
dynamics:
\begin{equation}
  R(E_1 \rightarrow E_2) = \left\{
\begin{array}{ll}
   R_0\exp\left[\displaystyle{\frac{-(E_2-E_1)}{T}}\right]
     &\mbox{for\ \ $E_2 > E_1$} \\
   R_0 &\mbox{for\ \ $E_2 < E_1$}.
\end{array}
\right.
\end{equation}
where $R(E_1 \rightarrow E_2)$ represents the transition rate for
a single polypeptide chain from state 1 to 2 with total energies
$E_1$ and $E_2$, respectively. $R_0$ is a overall constant
describing the inverse time scale for the transition process
between configurations (usually $R_0$ has the order of inverse
nanoseconds). Therefore the transition rate from one
conformational state to a neighboring state is determined by the
energy difference of these two states. Further analytic treatment
to this problem is made by utilizing the continuous time random
walk (CTRW)\cite{Montroll}. By this construction one is able to
reduce the multi-dimensional random walk problem to a
one-dimensional CTRW, resulting in a generalized master equation.
Schematically, one can first categorize the energy landscape by
the order parameter $\rho$, along which an energy distribution
function $P(E,\rho)$ is given (which is a Gaussian with mean and
width shown in Eqs.~(2) and (3)). With the use of Metropolis
dynamics one can calculate the associated transition rate
distribution function $P(R,\rho)$, specifying the jumping rate $R$
for a molecule at a state with order parameter $\rho$ to its
neighboring states, and therefore obtain the corresponding
waiting-time distribution $\Psi(\tau, \rho)$ for a molecule to
stay at a conformational state for time $\tau$ before it leaves. A
CTRW can be constructed by knowing both the waiting-time
distribution for the system and the jumping probabilities between
successive $\rho$'s. The latter is approximated to be
time-independent, which is equivalent to the quasi-equilibrium
assumption. By this assumption one can calculate these
probabilities utilizing the asymptotic distribution:
\begin{equation}
  \lim_{\tau\rightarrow\infty} G(\rho,\tau) \propto e^{-\beta F(\rho)},
\end{equation}
where $G(\rho,\tau)$ is the probability distribution function for
the polypeptide chain at time $\tau$, and $\beta=1/T$.

A generalized kinetic  master equation can thus be derived using
the CTRW approximation and conveniently represented in the
Laplace-transformed space as:
\begin{equation}
  sG(\rho,s)-n_0(\rho)=\hat{\mathcal{K}}(\rho,s)G(\rho,s)
\end{equation}
where $n_0(\rho)$ is the initial distribution of $G(\rho,\tau)$
and $\hat{\mathcal{K}}(\rho,s)$ is a linear operator which is
related to both the waiting time distribution and the jumping
probabilities. In the local connectivity case the generalized
master equation is reduced to a generalized (By generalized, we
mean instead of the usual Fokker-Planck equation where diffusion
is a constant in time representing a typical kinetic Markovian
behavior, here we obtain a non-Markovian diffusion kernel in time
due to the dimensional reduction from multiple one to a single
$\rho$) Fokker-Planck equation in the Laplace-transformed space:
\begin{equation}
  s\tilde{G}(\rho,s)-n_0(\rho)=\frac{\partial}{\partial\rho}\left\{D(\rho,s)
  \left[\tilde{G}(\rho,s)\frac{\partial}{\partial\rho}U(\rho,s)+
  \frac{\partial}{\partial\rho}\tilde{G}(\rho,s)\right]\right\},
\end{equation}
where
\begin{equation}
  U(\rho,s) \equiv \frac{F(\rho)}{T} + \log\frac{D(\rho,s)}{D(\rho,0)},
\end{equation}
$s$ is the Laplace transform variable over time $\tau$, and
$D(\rho,s)$ is the frequency-dependent diffusion parameter.
$F(\rho)$ is the average free energy derived from the random
energy model. The explicit expression for $D(\rho,s)$ is given in
the appendix. $s$, which has an unit of inverse time, is the
Laplace transform variable over time $\tau$. $\tilde{G}(\rho,s)$
is the Laplace transform of $G(\rho,\tau)$, which is the
probability density function such that $G(\rho,\tau)d\rho$ is the
probability for a protein to stay between $\rho$ and $\rho+d\rho$
at time $\tau$. Here $n_0(\rho)$ is the initial condition for
$G(\rho,\tau)$.

The boundary conditions for the generalized Fokker-Planck equation
are set as a reflecting one at $\rho=0$, where all the residues
are in their non-native states:
\[\left.\left[\tilde{G}(\rho,s)\frac{\partial}{\partial\rho}U(\rho,s)+
\frac{\partial}{\partial\rho}\tilde{G}(\rho,s)
\right]\right|_{\rho=0}=0,\] and an absorbing one at
$\rho=\rho_f$, where most of the residues are in the native
states:
\[\tilde{G}(\rho_f,s)=0.\]
The choice of an absorbing boundary condition at $ \rho=\rho_f $
facilitates our calculation for the first-passage time
distribution.

Alternatively, one can rewrite this generalized Fokker-Planck
equation in its integral equation form by integrating it twice
over $ \rho $:
\begin{eqnarray}
\tilde{G}(\rho,s) &=&
  -\int_{\rho}^{\rho_f}d\rho'\int_{0}^{\rho'}d\rho''\left[
  s\tilde{G}(\rho'',s)-n_0(\rho'')\right]\frac{\exp\left[U(\rho',s)-U(\rho,s)
  \right]}{D(\rho',s)} \nonumber \\
&=& -\int_{\rho}^{\rho_f}d\rho'\int_{0}^{\rho''}d\rho''\left[
  s\tilde{G}(\rho'',s)-n_0(\rho'')\right]\frac{K(\rho')}{D(\rho,s)K(\rho)}\, ,
\end{eqnarray}
where
\begin{equation}
  K(\rho) \equiv \frac{e^{\beta F(\rho)}}{D(\rho,0)}.
\end{equation}

In this paper, the first-passage time (FPT) to reach $ \rho_f $
(that is, the time required for the random walker to visit order
parameter value $ \rho_f $ for the first time) is used as a
typical or representative time scale for folding. One has the
following relation for the FPT distribution function
$P_{FPT}(\tau)$ :
\begin{equation}
  P_{FPT}(\tau) = \frac{d}{d\tau}(1-\Sigma) = - \frac{d\Sigma}{d\tau}
\end{equation}
where
\begin{equation}
  \Sigma(\tau) \equiv \int_{0}^{\rho_f} d\rho\, G(\rho,\tau)\, .
\end{equation}
The moments of the FPT distribution function are calculated from
the following relation:
\begin{eqnarray}
  \langle \tau^n \rangle &\equiv& \int_{0}^{\infty}d\tau\, \tau^n
  P_{FPT}(\tau)
  =-\int_{0}^{\infty}d\tau\, \tau^n\frac{d\Sigma(\tau)}{d\tau} \nonumber \\
  &=& n\int_{0}^{\infty}d\tau\, \tau^{n-1}\Sigma(\tau) = n\int_{0}^{\rho_f}d\rho
  \int_{0}^{\infty}d\tau\, \tau^{n-1}G(\rho,\tau) \nonumber \\
  &=& \bigg[ n(-1)^{n-1}\int_{0}^{\rho_f}d\rho\, \left(\frac{\partial}
  {\partial s}\right)^{n-1}\tilde{G}(\rho,s)
  \bigg]\bigg|_{s=0}.
\end{eqnarray}
If we make a series expansion of $\tilde{G}(\rho,s)$ and
$1/D(\rho,s)$:
\begin{equation}
  \tilde{G}(\rho,s) = \tilde{G}_0(\rho) + s \tilde{G}_1(\rho) + s^2
  \tilde{G}_2(\rho) + \cdots ,
\end{equation}
and
\begin{equation}
  1/D(\rho,s) = a_0(\rho) + s a_1(\rho) + s^2 a_2(\rho) + \cdots ,
\end{equation}
then we have
\begin{equation}
  \langle \tau^n \rangle = n! (-1)^{n-1}\int_{0}^{\rho_f}d\rho\,
  \tilde{G}_{n-1}(\rho),
  \nonumber
\end{equation}
and Eq.~(9) becomes
\begin{equation}
  \tilde{G}_{0}(\rho)=
  \int_{\rho}^{\rho_f}d\rho'\int_{0}^{\rho''}d\rho''
  n_0(\rho'')a_{0}(\rho) K(\rho,\rho')
\end{equation}
and for $n \geq 0$
\begin{equation}
  \tilde{G}_{n+1}(\rho)=
  -\int_{\rho}^{\rho_f}d\rho'\int_{0}^{\rho''}d\rho''\left[
  \sum_{j=0}^n
  \tilde{G}_{n-j}(\rho'')a_{j}(\rho)-n_0(\rho'')a_{n+1}(\rho)\right]K(\rho,\rho')
\end{equation}
by matching each coefficient of $s^n$ in Eq.~(16) and Eq.~(17).
Therefore one can calculate $\tilde{G}_n(\rho)$ recursively. Mean
while, one can also solve the integral equation (11) directly for
$\tilde{G}(\rho,s)$, and by the observation that
\begin{equation}
  \tilde{P}_{FPT}(s) = 1 - s \tilde{\Sigma}(s),
\end{equation}
where $\tilde{P}_{FPT}(s)$ and $\tilde{\Sigma}(s)$ are Laplace
transforms of $P_{FPT}(\tau)$ and $\Sigma(\tau)$, respectively, we
can investigate $P_{FPT}(\tau)$ by studying the behavior of
$\tilde{P}_{FPT}(s)$. To solve Eq.~(11) numerically, one first
replaces the integrations by discrete summations (recalling that
originally this model is itself constructed in a discrete
order-parameter space with equal spacings). Because Eq.~(11) is
linear in $\tilde{G}(\rho,s)$, one can solve for
$\tilde{G}(\rho,s)$ in the discrete $\rho$ space by a matrix
inversion technique. In the discrete version Eq.~(11) becomes
\begin{equation}
  \vec{G}_i=-\sum_{j, k}(\Delta\rho)^2 \hat{K}^{-1}_{ii}\hat{D}^{-1}_{ii}
  \hat{I_2}_{ij}\hat{K}_{jj}\hat{I_1}_{jk} (s\vec{G}_k-
  \vec{n_0}_k),
\end{equation}
where the integral operators become matrices
\begin{eqnarray}
\hat{I_1}_{ij}=\left\{
\begin{array}{cc}
1 &\mbox{\ \ \ \ \ \ if $i>j$,} \\
0 &\mbox{\ \ \ \ \ \ otherwise.}
\end{array}
\right. \nonumber \\
\hat{I_2}_{ij}=\left\{
\begin{array}{cc}
1 &\mbox{\ \ \ \ \ \ if $i\leq j$,} \\
0 &\mbox{\ \ \ \ \ \ otherwise}.
\end{array}
\right.
\end{eqnarray}
$i$, $j$, and $k$ are discrete labels of the order parameter.
$\hat{K}$ and $\hat{D}$ are diagonal matrices with nonzero
elements $K(\rho_i)$ and $D(\rho_i,s)$, respectively. $\vec{G}$
and $\vec{n_0}$ are vectors of elements $\tilde{G}(\rho_i,s)$ and
$n_0(\rho_i)$. With these notations one can easily get
\begin{equation}
  \vec{G}=(\hat{D}\hat{K}+s\hat{I_2}\hat{K}\hat{I_1})^{-1}
  \hat{I_2}\hat{K}\hat{I_1}\vec{n_0}
\end{equation}
In our calculations we make the matrix inversion with a Gaussian
elimination method with scaled-column pivoting, and the results
don't change very wildly with the number of grids we choose in the
discrete space, so we conclude this is a stable procedure for
solving $\tilde{G}(\rho,s)$ and therefore $\tilde{P}_{FPT}(s)$.

In summary, one first defines an order parameter $\rho$ describing
the direction of the folding process. Then one assigns the
distribution of energy states $P(E,\rho)$ along each $\rho$. The
next step is to specify the network, i.e., the geometry of the
energy landscape. In the present model this is done by random
connections between successive $\rho$. Finally, in order to
simplify the complex multi-dimensional random walk problem into an
one-dimensional problem, one utilizes the approach of continuous
time random walk (CTRW) and squeezes the vast information on the
energy landscape into the basic two requirements for constructing
a CTRW: a waiting-time distribution function and jumping
probabilities among different $\rho$'s. From the CTRW one gets a
generalized master equation, and in the local-connecting case it
simplifies into a generalized Fokker-Planck equation (Eq.~(9)).

One should note that the original Hamiltonian (Eq.~(1)) does not
suffice in building the whole energy landscape. In the approach
used here it is supplemented by the assumption of independence
among different energy states (The effects of correlations have
been discussed elsewhere\cite{PWW}.). The resulting energy
landscape has many features which may resemble real polypeptide
chains. First, there exists local energy traps, some of which are
very deep. Second, near the folded state the energy fluctuations
become much smaller, and diffusions close to the folded state are
fast (see next section). This implies that the folded state may
have some degrees of flexibility, allowing certain occupation of
substates and conformational changes. Finally, the random
connecting network (energy landscape) builds a strongly
interacting environment in which some effects of the cooperativity
may have already been reflected. The current approach is a
phenomenological one which captures the physics on the
coarse-grained or renormalized level. This is in analogy to the
Landau-Ginzburg approach for treating critical phenomena. A
microscopic Hamiltonian that is both computationally manageable
and quantitatively faithful to real folding is currently still not
available. The definition of the order parameter $\rho$ needs to
be looked at more carefully, since there might be ambiguities in
defining the fraction of native conformations in real proteins.
However, parameters like $\delta \epsilon$ and $\Delta \epsilon$
appear to characterize the bias and the fluctuation on the energy
landscape faithfully. Hence, one can hope to build up a model
which has several important features resembling real proteins in a
semiquantitative way, and it is interesting to see how far one can
go with this approach.

\section{Results and Discussions}
We start the numerical calculations of moments and distributions
of FPT by setting $R_0 = 10^9 s^{-1}$, $N = 100$ and $\nu = 10$.
For simplicity we assume $\delta\epsilon = \delta L$ and
$\Delta\epsilon = \Delta L$. The ratio of the energy gap between
the native state and the average non-native states versus the
fluctuations of the non-native states,
$\delta\epsilon/\Delta\epsilon$, which represents the relative
degree of the funnel slope or the bias towards the native state
compared with the local roughness or the traps of the folding
landscape, becomes an adequate parameter for this model. We set
the initial distribution of the protein molecules to be $n_0(\rho)
= \delta(\rho-\rho_i)$, where $\rho_i$ is set to be 0.05. In our
calculations we set $\rho_f = 0.9$, which means that 90 percent of
the amino acids are in their native states. We calculate the
moments $\langle \tau^n \rangle$ of the FPT distribution function
by utilizing Eqs.~(11)-(21).

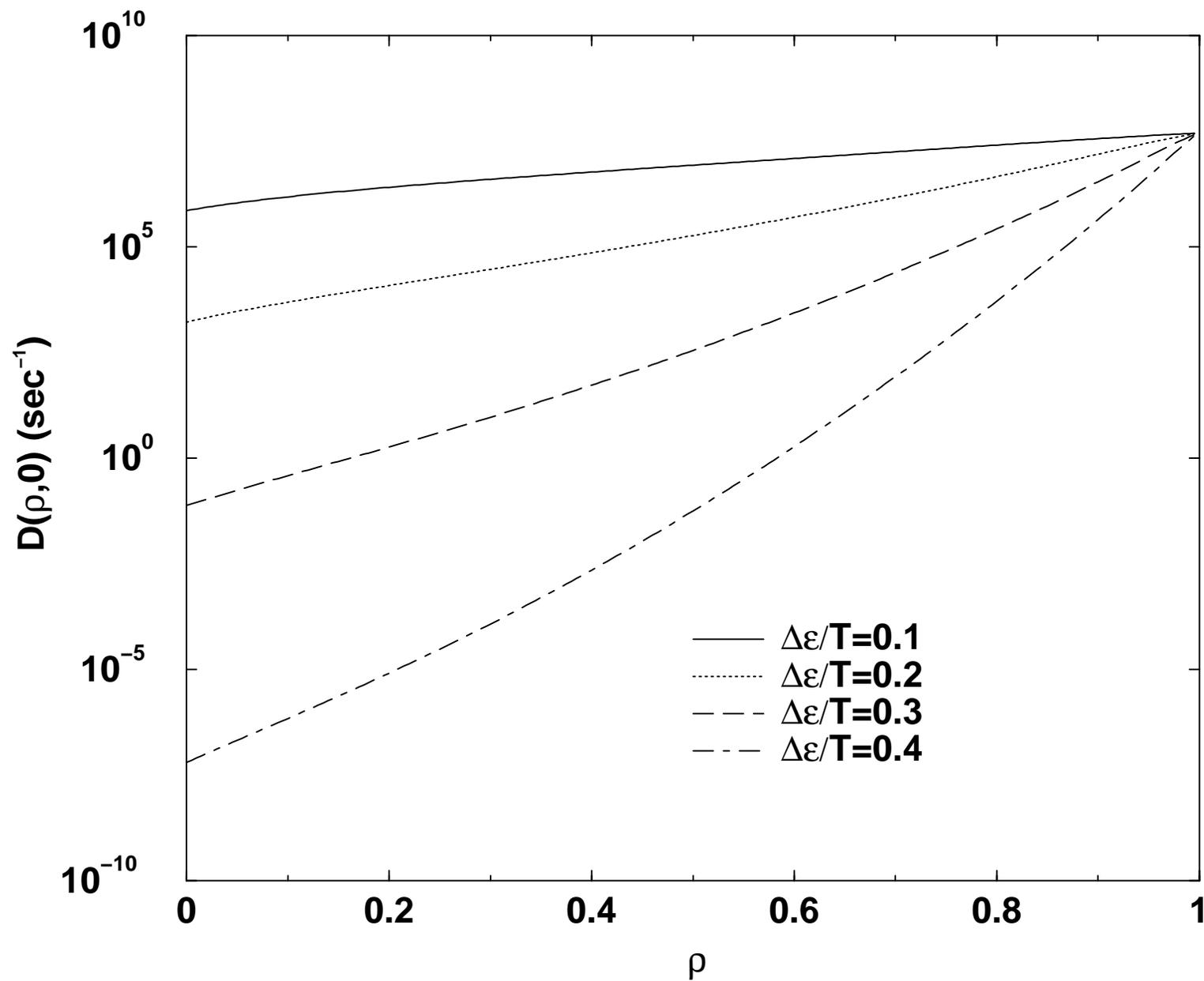
\begin{figure}[Fig1]
\caption{$D(\rho,0)$ for various settings of $\Delta \epsilon/T$.
At low temperature the diffusion timescale varies in orders of
magnitude over $\rho$.}
\label{diff}
\end{figure}

Fig.~\ref{diff} shows an example of $D(\rho,s=0)$ for various
settings of $\Delta \epsilon/T$. One can see clearly that the
diffusion is very fast near $\rho_f$. This means that usually
the conformation changes near the folded state don't play a major
role in the folding time. It also indicates the flexibility in
conformational changes near the folded state. At higher
temperature (hence smaller $\Delta \epsilon/T$) the diffusion
parameter for different $\rho$'s has less distinction, whereas at
low temperature the diffusion process varies in large orders of
magnitude with $\rho$.

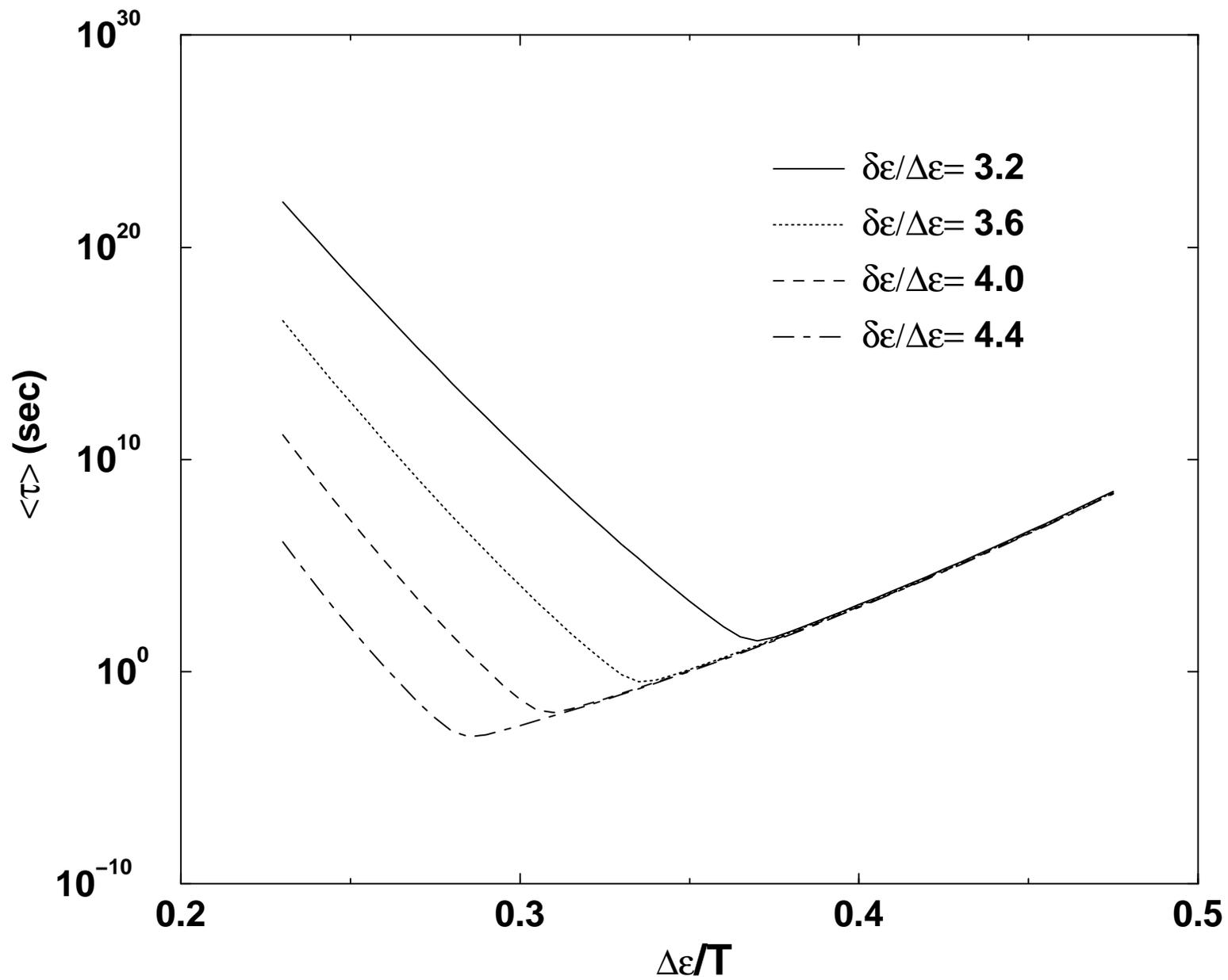
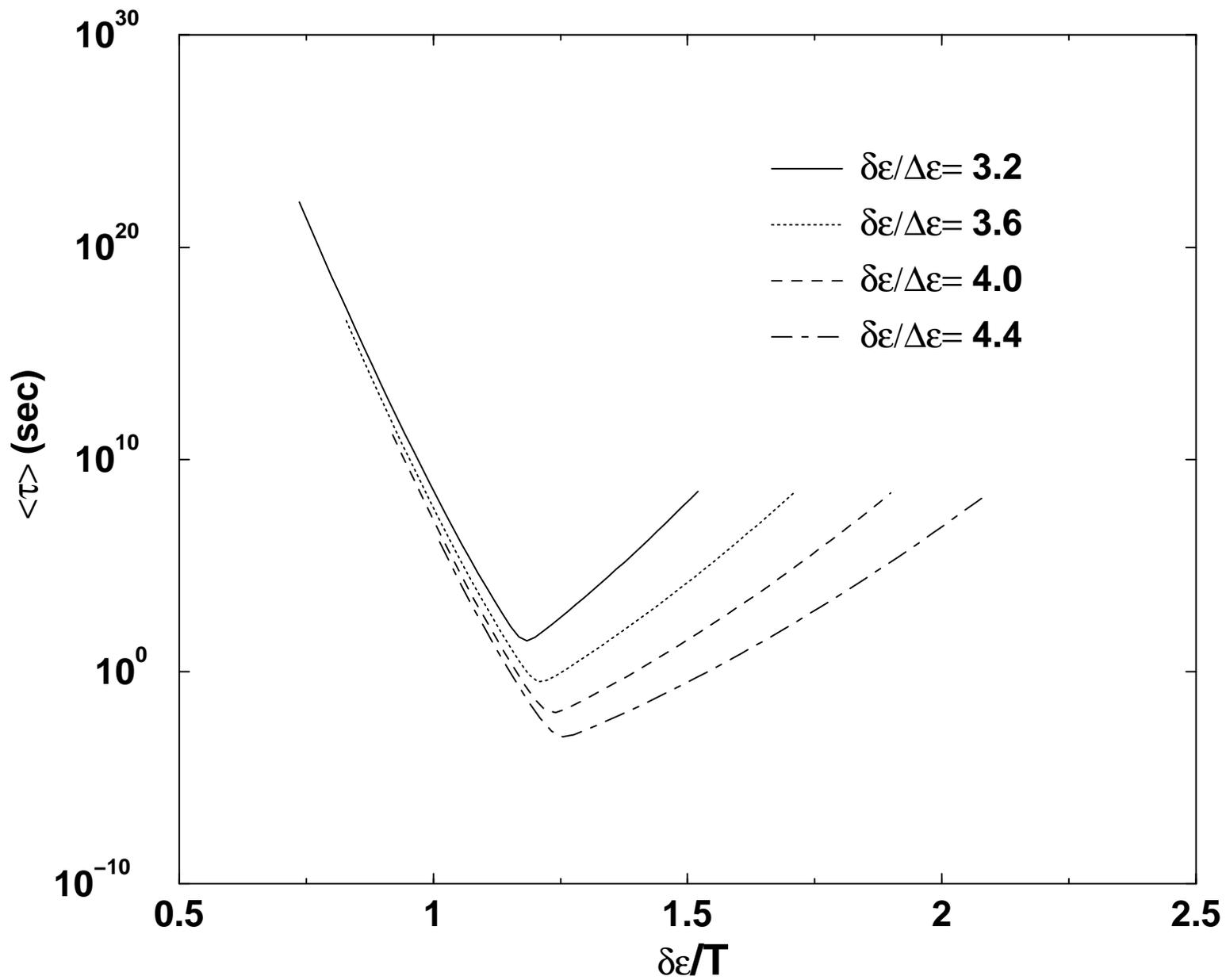
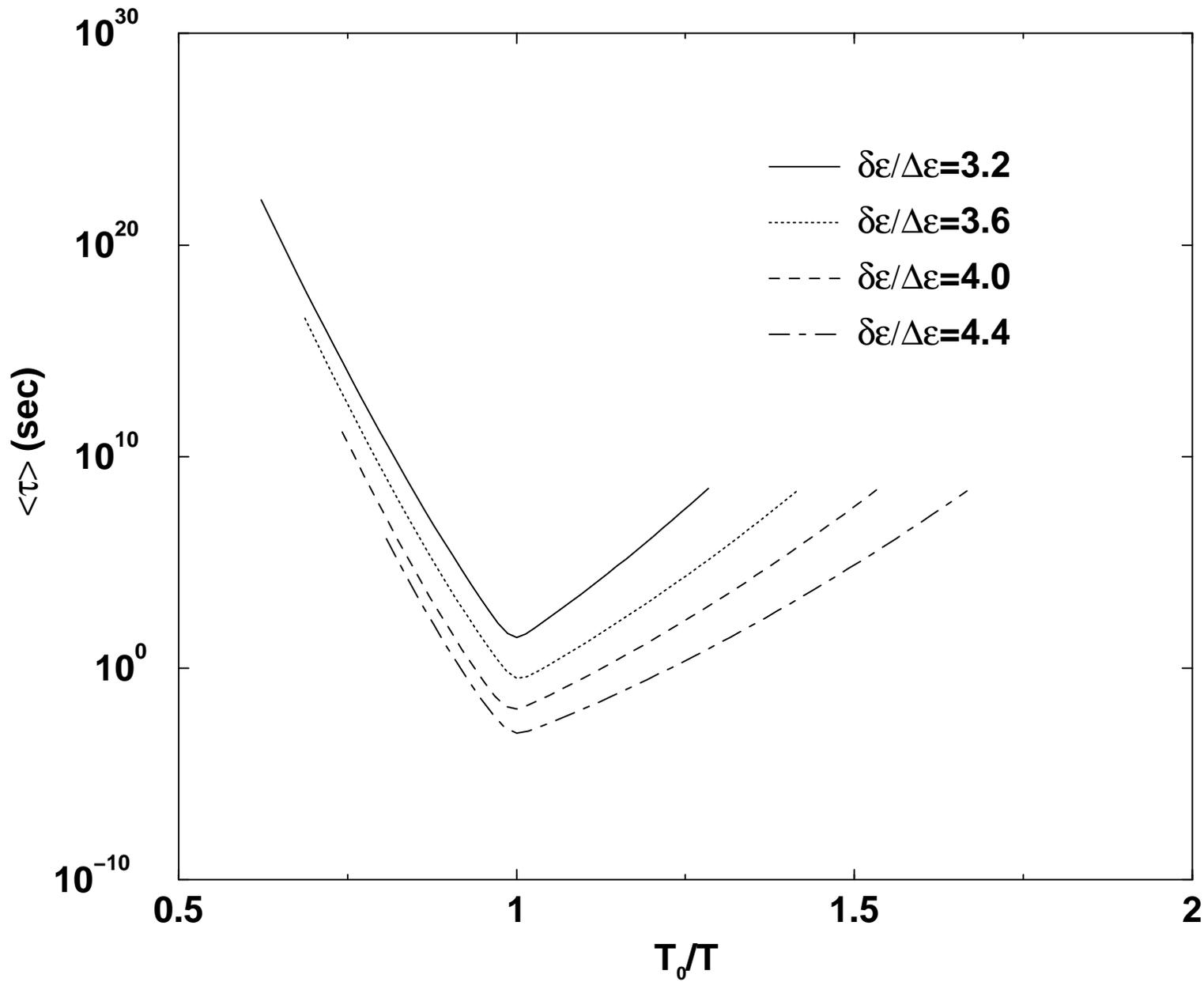
\begin{figure}[Fig2]
\caption{MFPT versus different scaled inverse temperatures ((a):
$\Delta\epsilon/T$; (b): $\delta\epsilon/T$; (c):$T_0/T$) for
various $\delta\epsilon/\Delta\epsilon$. For each fixed
$\delta\epsilon/\Delta\epsilon$, the curve is like an V shape. The
temperature at which MFPT is at its bottom is defined as the
transition temperature $T_0$. As $\delta\epsilon/\Delta\epsilon$
increases, the minimum of MFPT decreases.}
\label{t1}
\end{figure}

The MFPT $\langle \tau\rangle$ for the folding process versus
different scaled inverse temperatures are plotted in Fig.~\ref{t1}
for various settings of the parameter
$\delta\epsilon/\Delta\epsilon$. Note that the vertical axis is in
the logarithmic scale. We have a letter V curve for each fixed
$\delta\epsilon/\Delta\epsilon$. At high temperature, the MFPT is
large although the diffusion process itself is fast (i.e.,
$D(\rho,s)$ is large). This long-time folding behavior is due to
the instability of the folded state. The MFPT is also large at low
temperature. This is due to the importance of the onset of the low
energy non-native trapped states.

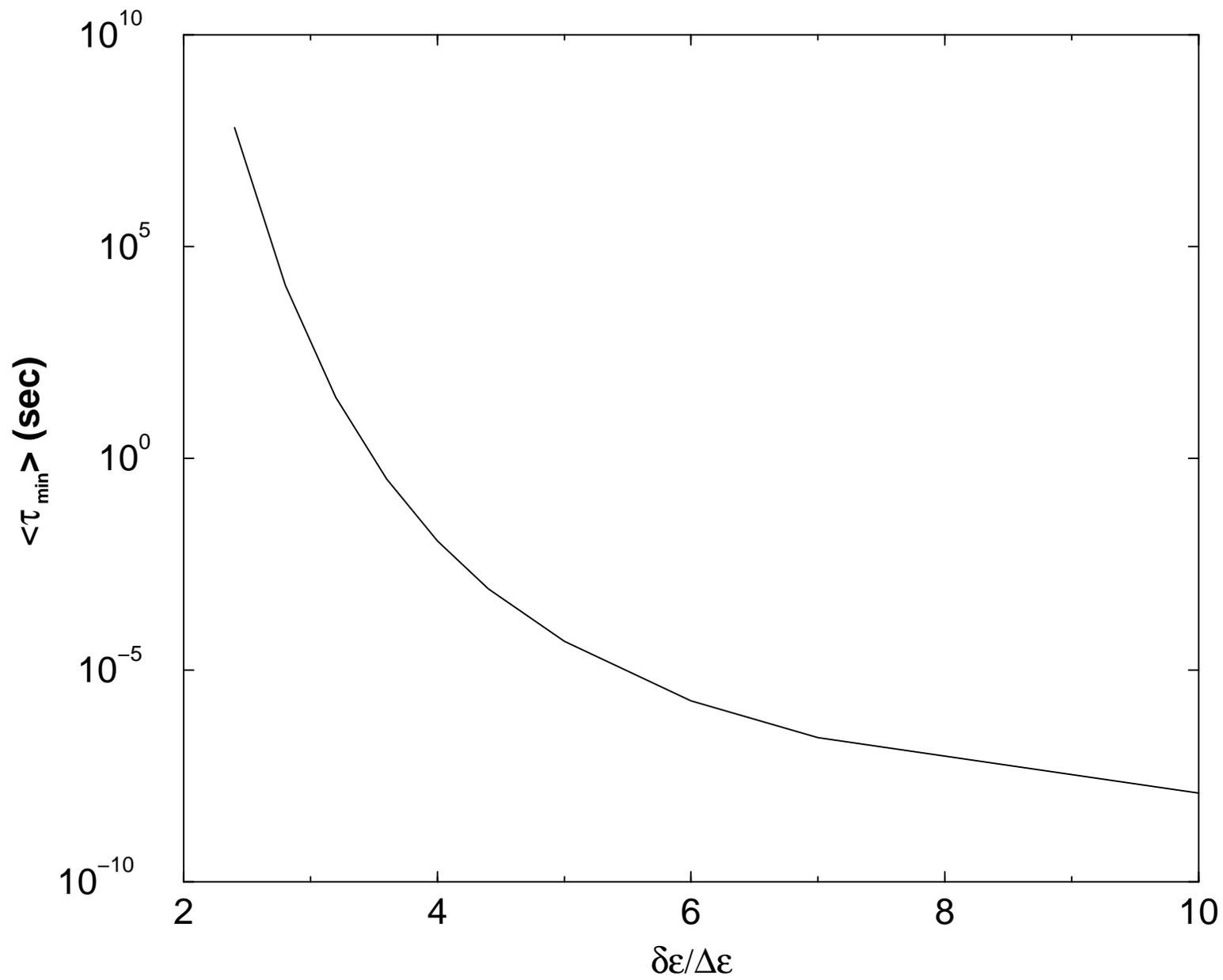
\begin{figure}[Fig3]
\caption{The minimum of MFPT $\langle \tau\rangle_{min}$ in a
logarithmic scale versus $\delta\epsilon/\Delta\epsilon$. There
exists a monotonic relationship between $\langle \tau
\rangle_{min}$ and $\delta\epsilon/\Delta\epsilon$. As
$\delta\epsilon/\Delta\epsilon$ increases, the minimum of MFPT
monotonically decreases.}
\label{t_min}
\end{figure}

The MFPT reaches its minimum at a transition temperature $T_0$. By
comparing this minimum for various values of $\delta\epsilon/
\Delta\epsilon$, we find that the minimum of MFPT gets smaller by
increasing the ratio of energy bias versus roughness (see
Fig.~\ref{t_min}). This indicates that a possible criterion to
select the subset (subspace) of the whole sequence space leading
to well designed fast folding protein is the maximization of
$\delta\epsilon/\Delta\epsilon$. In other words, one has to choose
the sequence subspace such that the global bias overwhelms the
roughness of the energy landscape\cite{Goldstein}.

By comparing the MFPT curves for various set of $\delta\epsilon/
\Delta\epsilon$, we found that when $T<T_0$, the MFPT is only
dependent on $\Delta\epsilon/T$. This indicates that the roughness
condition dominates in the MFPT for this temperature regime. On
the other hand, when $T>T_0$, the MFPT is mainly dependent on
$\delta\epsilon/T$ and has less dependence on $\Delta\epsilon/T$.
This suggests that the transition at $T=T_0$ marks the crossover
between competing contributions due to $\Delta\epsilon/T$ and
$\delta\epsilon/T$.

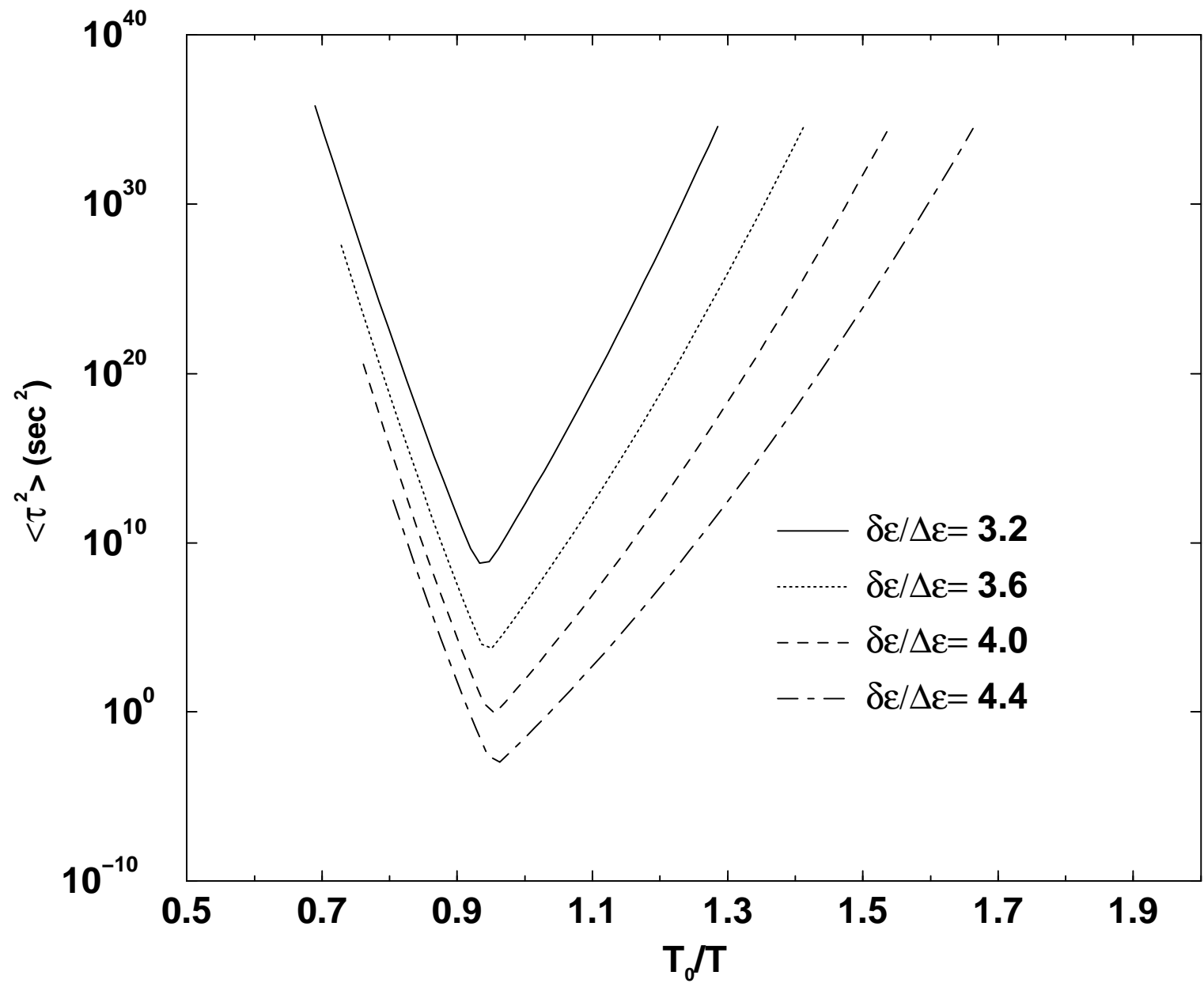
\begin{figure}[Fig4]
\caption{ $\langle \tau^2 \rangle $ versus reduced inverse
temperature $\Delta\epsilon/T$ for various
$\delta\epsilon/\Delta\epsilon$. For each fixed
$\delta\epsilon/\Delta\epsilon$, the curve is like an V shape. The
temperature at which MFPT is at its bottom is defined as the
transition temperature $T_0$. As $\delta\epsilon/\Delta\epsilon$
increases, the minimum of $ \langle \tau^2 \rangle $ decreases.}
\label{t2}
\end{figure}

We also calculate the higher moments of the FPT distribution. In
Fig.~\ref{t2} we present the results for $\langle \tau^2\rangle$
as an illustrated example. Again we have a V shape curve for each
setting of $\delta\epsilon/ \Delta\epsilon$ with the minimum of
the curve having a temperature close to $T_0$. Similar to the
behavior of $\langle \tau\rangle$, at low temperature $\langle
\tau^2\rangle$ is only dependent on $\Delta \epsilon/T$, and in
the high temperature regime it is mainly dependent on $\delta
\epsilon/T$. For higher moments we also obtain the same
conclusion.

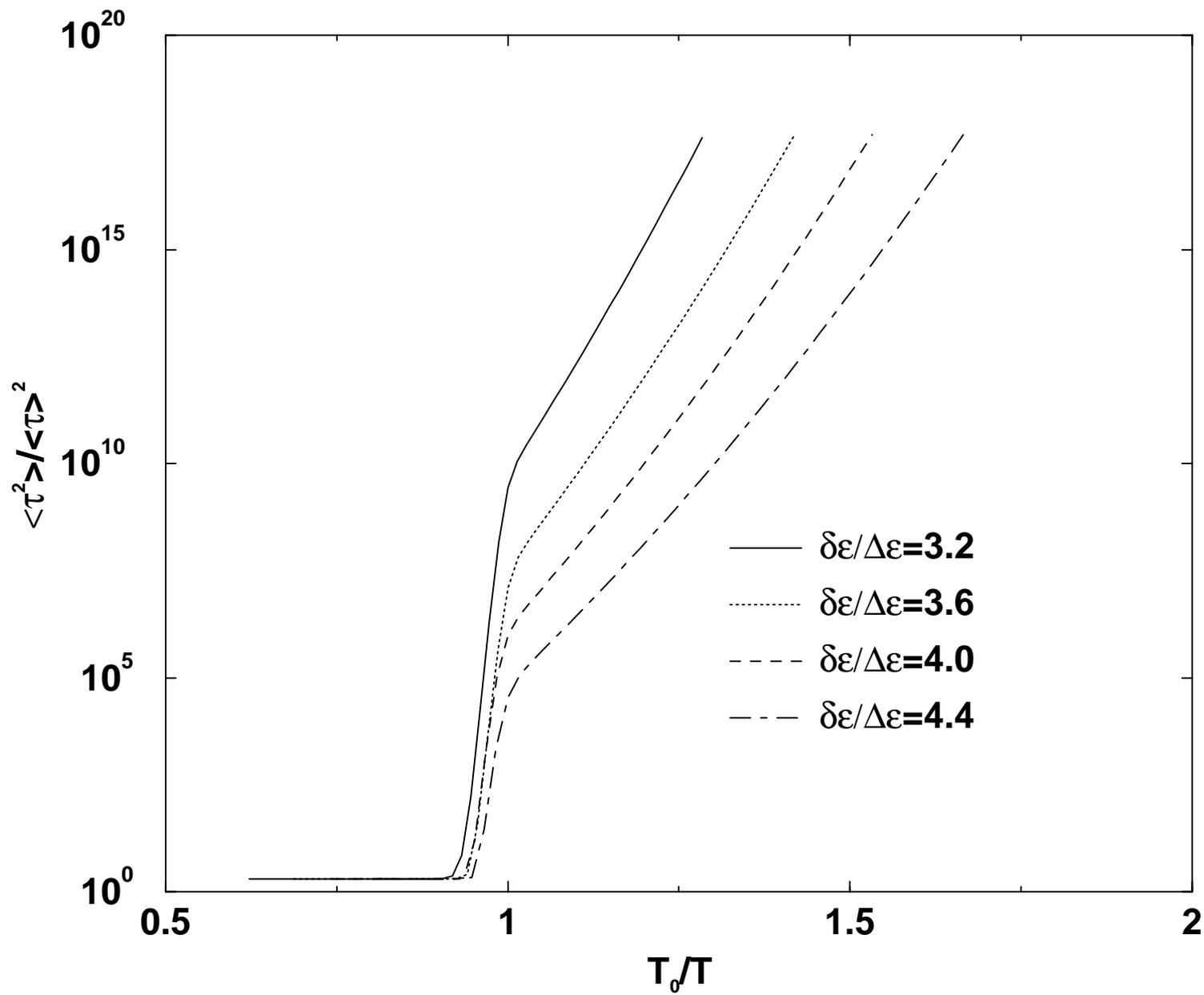
\begin{figure}[Fig5]
\caption{$\langle \tau^2\rangle/ \langle \tau \rangle^2$
  versus reduced inverse temperature $\Delta\epsilon/T$ for various
  $\delta\epsilon/\Delta\epsilon$. At high temperature this value keeps
  finite and the folding process is self-averaging. As the temperature drops,
  the value starts to diverge and non-self-averaging behavior emerges. }
\label{reduced_t2}
\end{figure}

In Fig.~\ref{reduced_t2} we show the behavior of the reduced
second moment, $\langle \tau^2\rangle/ \langle \tau\rangle^2$. We
find that the reduced second moment starts to diverge at
temperature around $T_0$, where MFPT is at its minimum. The degree
of divergence increases rapidly as temperature drops below $T_0$.
This indicates that the average is not a good representative of
the system any more and a long tail in the FPT distribution is
developed. The intermittency where rare events make great
contribution occurs. The divergence of the second moment also
shows that the dynamics is exhibiting non-self-averaging behavior.

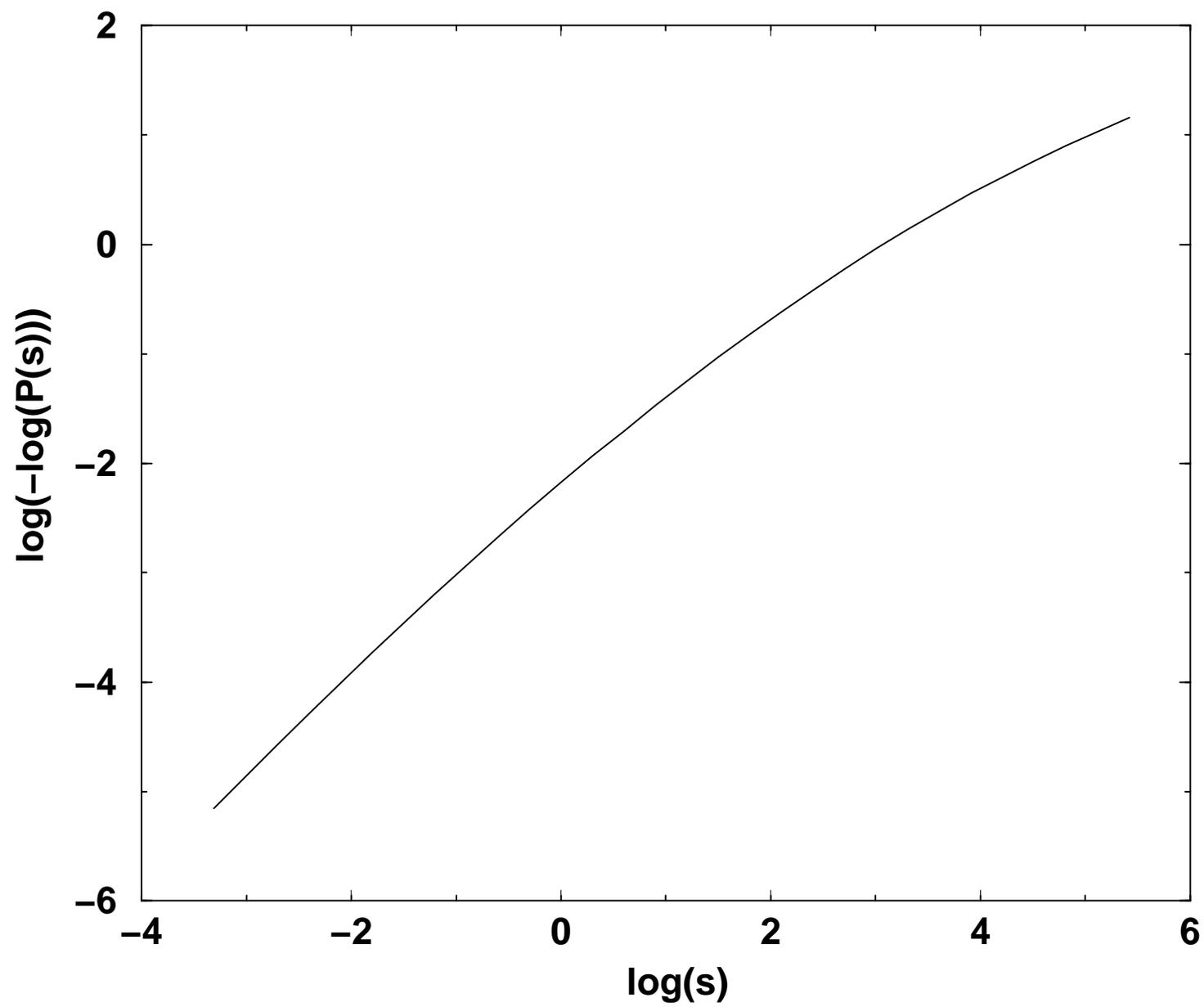
\begin{figure}[Fig6]
\caption{An illustrative figure of the power law distribution as
compared with a Gaussian normal distribution. Notice the prominent
fatty tails of power law distribution. }
\label{stretch_exp}
\end{figure}

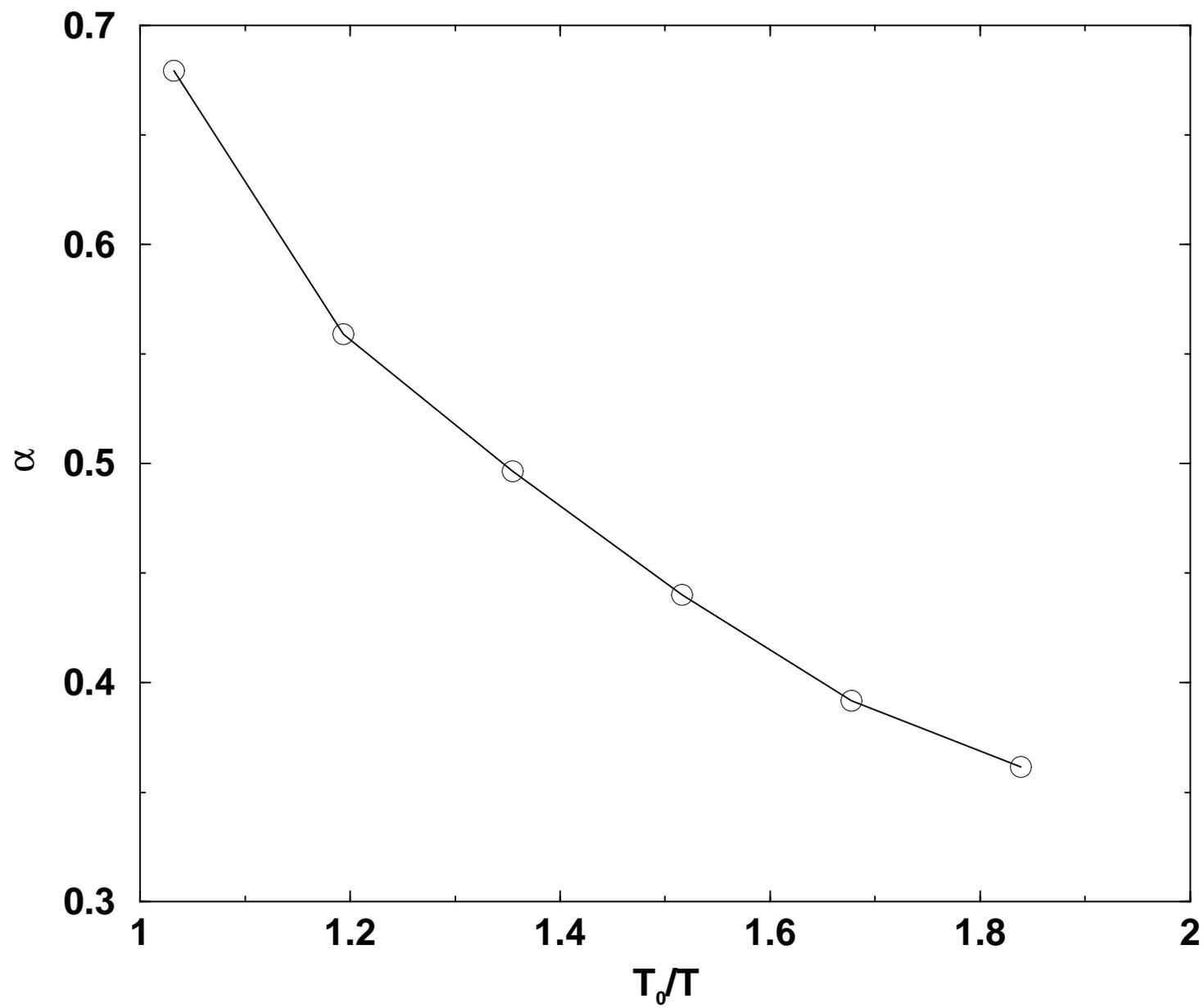
\begin{figure}[Fig7]
\caption{The exponent $\alpha$ versus $\Delta\epsilon/T$ for the
case $\delta\epsilon/\Delta\epsilon = 4.0$. Below the transition
temperature $T_0$, the FPT distribution is close to a L\'evy
distribution, and for large time it has a power-law tail :
$P_{FPT}(\tau) \sim \tau^{-(1+\alpha)}$. We see that below $T_0$,
$\alpha$ decreases when the temperature gets lower.}
\label{exponent}
\end{figure}

>From the study of higher moments, we find at high temperature the
relationship $\langle \tau^n\rangle = n!\langle \tau\rangle^n$.
Therefore the FPT distribution function is Poissonian in the high
temperature regime. But when $T < T_0$, it is hard to get more
information from the moments because of their diverging behavior.
On the other hand, the folding dynamics can be also studied by
solving the linear integral equation (11) directly by making the
inversion of the linear operator. We investigate the behavior of
the FPT distribution function in the Laplace-transformed space
derived via Eqs.~(12) and (19). Our result shows that for $T <
T_0$ $\tilde{P}_{FPT}(s)$ decays slowly over decades, which
suggests that the usual numerical Laplace inversion techniques
cannot be applied. Moreover, if we plot $\log(-\log P(s))$ versus
$\log s$, we find that there is approximately a linear relation
over several orders of magnitude in $s$ (see
Fig.~\ref{stretch_exp} for example). This indicates that for $T <
T_0$ $\tilde{P}_{FPT}(s)$ can be approximated by a stretched
exponential:
\begin{equation}
 \tilde{P}_{FPT}(s) \approx e^{-c s^{\alpha}} \, ,
\end{equation}
which is the Laplace transform of the L\'evy distribution in the
time space. So we have
\begin{equation}
  P_{FPT}(\tau) \approx -\frac{1}{\pi}\sum_{n=1}^{\infty}
  \frac{(-c)^n}{\tau^{\alpha n+1}} \frac{\Gamma(\alpha n+1)}{\Gamma(n+1)}
  \sin (\pi\alpha n) \, .
\end{equation}
$\alpha$ lies between 0 and 1. From the asymptotic property of the
L\'evy distribution function we learn that $P_{FPT}(\tau) \sim
\tau^{-(1+\alpha)}$ for large $\tau$. In Fig.~\ref{exponent} we
make a plot of the exponent $\alpha$ versus $\Delta\epsilon/T$ for
the case $\delta\epsilon/\Delta\epsilon = 4.0$. We find that
$\alpha$ decreases when the temperature decreases.

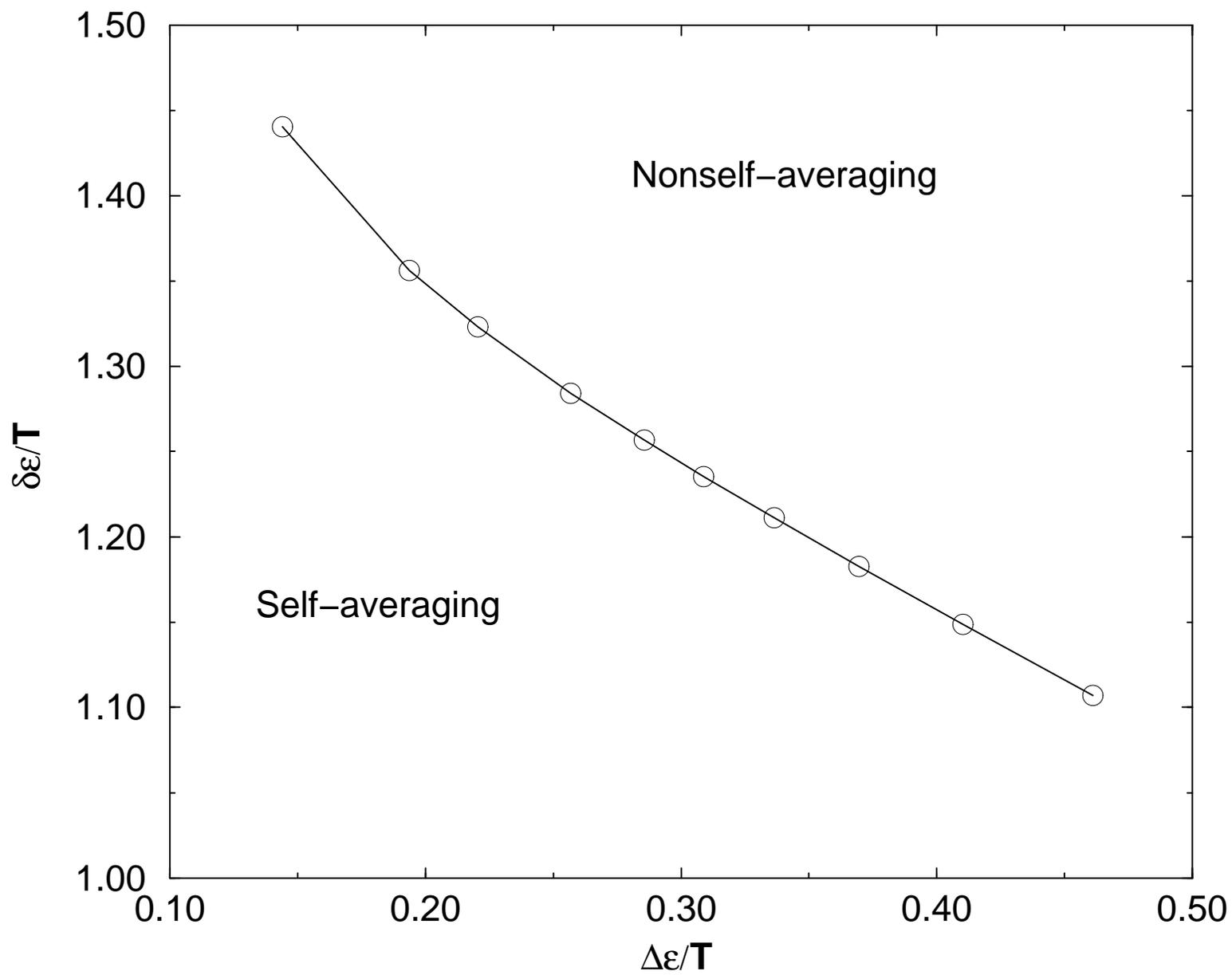
\begin{figure}[Fig8]
\caption{Dynamic phase diagram showing the transition between
self- and non-self-averaging dynamics. }
\label{phase_diagram}
\end{figure}

From the results above, we find that for a fixed energy landscape,
there exists a dynamic transition temperature $T_0$. When the
temperature is lager than $T_0$, the FPT distribution is
Possionian, which means exponential kinetics. In this case the the
underlined energy landscape of the polypeptide chain is rather
smooth, and therefore the folding process just follows the valleys
and saddle points on the landscape with minor fluctuations. On the
other hand, when the temperature is below $T_0$, the second and
higher reduced FPT moments diverge, and the FPT distribution
exhibits a power-law decay behavior. This indicates that the
folding process is non-self-averaging and achieved through
numerous timescales. In this case the underlined energy landscape
of the polypeptide chain is rather rough. Nearby folding paths on
the energy landscape may have big differences in their energy
barriers. In Fig.~\ref{phase_diagram} we make a "phase diagram"
showing this dynamical character. We found that for a fixed
$\delta\epsilon/\Delta\epsilon$ (which corresponds to a straight
line through the origin in the phase diagram), when the
temperature is lowered, the system goes through a transition from
self-averaging to non-self-averaging behavior. Furthermore, for a
fixed $\delta\epsilon/T$, the dynamics becomes non-self-averaging
after we increase $\Delta\epsilon/T$ over some critical value.
This can be accounted intuitively since when one increases
$\Delta\epsilon/T$, roughness on the energy landscape becomes more
prominent, and different folding paths will lead to very distinct
folding times. It is less intuitive however to see that if we fix
the control parameter $\Delta\epsilon/T$ and increase
$\delta\epsilon/T$, the dynamics also becomes non-self-averaging.
This can be understood by viewing that the MFPT is decreasing if
we keep increasing $\delta\epsilon/T$ until a transition point at
which it losses its dominance. When $\delta\epsilon/T$ is well
below its transition point, the dynamics is primarily dominated by
the effects of the relative bias ($\delta\epsilon/T$) and is
therefore self-averaging. When $\delta\epsilon/T$ is larger beyond
this transition point, the FPT distribution depends mainly on the
relative roughness ($\Delta\epsilon/T$) and the dynamics becomes
non-self-averaging. The results will not be much different if we
keep on increasing $\delta\epsilon/T$. Finally, from our results,
one sees that the non-self-averaging phenomena do not necessarily
indicate slow dynamics. For example, when one fixes
$\Delta\epsilon/T$ and increases $\delta\epsilon/T$, the dynamics
eventually becomes non-self-averaging. However, the MFPT keeps
decreasing until it reaches the transition point, after which the
MFPT keeps approximately constant.

For comparison, we also calculate the folding transition
temperature $T_f$ by solving the corresponding unfolding problem.
Specifically, we start from $\rho_i=0.9$ and calculate the
first-passage time to $\rho_f=0.1$. The unfolding MFPT curve
intersects the folding MFPT one on its high-temperature branch.
Since one expects the inverse of MFPT to characterize the rate of
the kinetic process, this point of intersection can reasonably be
used to locate the folding transition, where the kinetic rates of
the folding and unfolding processes are equal. The comparison of
this folding transition temperature $T_f$ with the dynamical
transition temperature $T_0$ is listed in Table 1. We find that in
all our examples $T_0$ is strongly correlated with $T_f$, which is
in agreement with the experimental facts and simulation
results\cite{Chan}.

As discussed above, this dynamical transition results from the
competition between different controlling parameters. In our
calculations, we only have two such terms: $\delta \epsilon$ and
$\Delta \epsilon$, and the result is a sharp crossover around
$T_0$. One should note that in general $\delta \epsilon \neq
\delta L$ and $\Delta \epsilon \neq \Delta L$. Furthermore, one
has to take into consideration the three- and higher-body
interactions among amino acid residues, which are also necessary
for accounting the cooperative behavior in folding dynamics. In
reality one would expect the sharp crossover transition to be
smoothed out if one take into account a more comprehensive set of
controlling parameters.

In single-molecule folding experiments, it is now possible to
measure not only the mean but also the fluctuations and moments as
well as the distribution of folding times\cite{Chu}. In different
experimental and sequence conditions, one can see different
behaviors of the folding time and its distributions. A well
designed fast-folding sequence with suitable experimental
condition exhibits self averaging and simple rate behavior.
Multiple routes are parallel and lead to folding. A less well
designed sequence will not only fold slowly but also often exhibit
non-self-averaging nonexponential rate behavior, indicating the
existence of intermediate states and local traps. In this case,
discrete paths to the folded state emerge. The folding process is
sensitive to which kinetic path it takes. Certain specific kinetic
paths give the crucial contribution to folding (rare events
contribute most, indicating intermittency). One can use
single-molecule experiments to uniquely determine the fundamental
mechanisms and intrinsic features of the protein folding. In
typical bulk experiments with large populations, it is very hard
to distinguish whether the non-exponential behavior is intrinsic
for each individual molecule or just the result of inhomogeneous
averaging of the large sample of molecules with each individual
molecule exhibiting single exponential behavior.

It is worth mentioning that in this paper although we focus on the
study of the protein folding problem, the approach we use is
generally applicable to problems with barrier crossing on a
multi-dimensional complex energy landscape. The main ingredient
for this model is Brownian motion on a rough multi-dimensional
landscape, or equivalently, a random walk on a complex network
with frustrated (highly fluctuating) environment. Since this is
quite general and universal, we expect our results may also be
able to account for a large class of phenomena. In fact the
experiments on glasses, spin glasses, viscous
liquids\cite{spinglass} and conformational dynamics already show
the existence of non-exponential distribution at low temperatures.
In particular, a recent experiment on single-molecule enzymatic
dynamics\cite{Xie} shows explicitly the L\'evy like distribution
of the fluctuation time for the underlined complex protein energy
landscape. A theoretical investigation of the complex-system
dynamics also showing the L\'evy like behavior using fractal
diffusion approach has recently been carried out\cite{barkai}. Our
approach provides a "microscopic" foundation and rationale for the
fractal model.

\section{Conclusion}
In this report, we have studied the kinetics of folding along a
locally-connected order-parameter path. We used a generalized
Fokker-Planck diffusion equation for describing the folding
dynamics. We further solved the equation and obtained the
expression for the MFPT. The effects of a stability gap $\delta
\epsilon$, temperature, and fluctuations of the folding landscape
on FPT were discussed. We found that the MFPT is smaller when the
ratio of the energy gap between the native and average non-native
states versus the fluctuations of the landscape, $\delta\epsilon /
\Delta\epsilon$ is larger.

The fluctuations and higher-order moments were also studied. It
was found that for temperatures well above $T_0$, the folding
process is self-averaging and its FPT distribution obeys a Poisson
distribution. But when the temperature is lower, the fluctuations
start to diverge. This means that the actual folding process may
happen on multiple time scales, and the non-self-averaging
behavior emerges. In this case, the full distribution of the FPT
is required in order to characterize the system. From our analysis
the distribution of FPT turns out to be close to a L\'evy
distribution, which has a power-law tail for long time. One
expects to be able to see this kind of fluctuations in
single-molecule experiments. In the bulk measurements the average
fluctuations are reduced due to the central limit theorem. Further
more one cannot tell if the fluctuations are either from the
intrinsic properties of the molecules or the inhomogeneous
averages over molecules. Our analytical results of the MFPT and
its self- or non-self- averaging behavior may provide a possible
kinetic basis for the criterion in selecting a subset of the whole
sequence space to reach well-designed and fast-folding proteins.

By carefully reexamining this model one finds that there are still
open questions for this whole set of constructions. First, it is
not clear whether the order parameter $\rho$ is the best reaction
coordinate for folding. Second, correlations between different
states are ignored. Third, in the model it is assumed that each
conformational state has $N\nu+1$ neighboring states, and the
number of total conformal states is equal to $(\nu+1)^N$. These
might be overestimated. In the real case, due to steric
restrictions, the number of available conformal states and their
connections might be greatly reduced. Nevertheless, from the
results in this paper, we see that this phenomenological model
indeed characterizes many features and provides several important
insights into the protein-folding problem.

\section*{Appendix}
In this appendix we give an explicit expression for the
frequency-dependent diffusion parameter $D(\rho,s)$, where s is
the Laplace transform variable over time $\tau$:
\begin{equation}
  D(\rho,s) \equiv \left(\frac{\lambda(\rho)}{2N^2}\right)\left\langle
  \frac{R}{R+s}\right\rangle_R (\rho) \Bigg/ \left\langle
  \frac{1}{R+s}\right\rangle_R (\rho)\, .
\label{diff_coeff}
\end{equation}
The average $\langle \ \rangle_R$ is taken over $P(R,\rho)$, the
probability distribution function of the transition rate $R$ from
one state with order parameter $\rho$ to its neighboring states,
which may have order parameters equal to $\rho-1/N$, $\rho$ or
$\rho+1/N$, and $\lambda(\rho) \equiv 1/\nu + (1 - 1/\nu)\rho $ is
the probability for a molecule to move to a state with
$\rho=\rho_0+1/N$ or $\rho=\rho_0-1/N$ when it leaves a state with
$\rho=\rho_0$. The probability distribution $P(R,\rho)$ is one of
the main ingredient for characterizing the dynamical behavior of
this protein folding model. For example, if the energy landscape
is smooth, the transition rates are distributed narrowly and can
be described by a single scale. Then we have the usual Markovian
diffusion. As an illustrated example, if one sets $P(R,\rho)
\approx \delta(R-R_a(\rho))$ for some specific $R_a(\rho)$, then
$D(\rho,s) \approx (\lambda(\rho)R_a(\rho))/ (2N^2)$, which is
independent of $s$. Hence Eq.~(9) just reduces to the normal
Fokker-Planck equation, and the folding process can therefore be
well described by a single time scale. On the other hand, if the
distribution $P(R,\rho)$ is broad over $R$, then the dynamic
behavior can not be discussed by using a single time scale. This
results a generalized Fokker-Planck equation with the diffusion
kernel in time. In this case one would expect a broader
distribution in the FPT distribution $P_{FPT}(\tau)$.

To derive the diffusion kernel, one first notice that the kinetic
rate  of jumping from one configuration to another is a stochastic
variable
\[R = R_0\sum_{E_i>E}\exp(-\frac{E_i-E}{T}) + R_0\sum_{E_i<E}1.\]
where $R$ is a function of a stochastic variable energy $E$.
Therefore by knowing the distribution of $E$, one can derive the
probability distribution of $R$ as well. The explicit expression
for $P(R,\rho)$ has been derived in the literature\cite{BW}, and
we just quote the result here:
\begin{equation}
  P(R,\rho) = \zeta\left(\frac{1}{\sqrt{2} \beta \Delta E(\rho)}
  \right)\delta(R-R_0 N\nu)
\end{equation}
for $R = R_0 N \nu$,
\begin{equation}
  P(R,\rho) = \frac{\displaystyle{\beta\Delta(\rho)\left[ 2\log\left(
  \frac{R_0 N\nu}{R}\right)\right]^{1/2} -\frac 12 \beta^2 \Delta
  E^2(\rho)}}{\displaystyle{R_0 N\nu\left[4\pi \log\left(\frac{R_0 N\nu}
  {R} \right)\right]^{1/2}}}
\end{equation}
for $R_0 N \nu > R > R_{sep}(\rho) $, and
\begin{equation}
  P(R,\rho) = \frac{1}{\sqrt{2\pi \beta^2 \Delta E^2(\rho) R^2}} \exp\left\{
  -\frac{\log^2(R/R_{sep}(\rho))}
  {2\beta^2\Delta E^2(\rho)}\right\}
\end{equation}
for $R_{sep}(\rho) \geq R \geq R_{slow}(\rho) .$ $P(R,\rho) = 0$
for $R<R_{sloq}$ and $R>R_0 N\nu$. Here $\beta$ is the inverse
temperature, $R_{sep}(\rho) \equiv R_0 N \nu \exp(- \beta^2 \Delta
E^2/2 ) $, and $R_{slow}(\rho) \equiv R_0 N\nu \exp[\beta^2\Delta
E^2(\rho)/2 - \sqrt{2S^*}\beta \Delta E(\rho)]$, where
\begin{equation}
  S^* \equiv N\left[-\rho\log(1-\rho)-(1-\rho)\log\frac{1-\rho}{\nu}\right],
\end{equation}
which is the configuration entropy for $N\rho$ native residues.
The function $\zeta(x)$ is defined as
\begin{equation}
  \zeta(x) \equiv \frac{1}{\sqrt{\pi}}\int_{x}^{\infty}\! dy
  \,e^{-y^2} \, ,
\end{equation}
which is equal to the complement error function $\mbox{erfc}(x)$
divided by $2$ for $x>0$.

From Eq.~(\ref{diff_coeff}), one can make a series expansion of
$1/D(\rho,s)$ with respect to $s$ as in Eq.~(17), which
facilitates the calculation for the moments of the FPT
distribution.. Here we just list the first several terms for
$a_n(\rho)$:
\begin{eqnarray}
  a_0(\rho)&=&\frac{2N^2}{\lambda(\rho)}R_1(\rho) \nonumber \\
  a_1(\rho)&=&\frac{2N^2}{\lambda(\rho)}(R_2(\rho)-R_1^2(\rho))
  \nonumber \\
  a_2(\rho)&=&\frac{2N^2}{\lambda(\rho)}(R_3(\rho)-2R_2(\rho)R_1(\rho)+R_1^3(\rho))
\end{eqnarray}
and etc., where $R_n(\rho) \equiv \langle 1/R^n\rangle_R(\rho)$.
In our calculations we carry out the average over rate
distribution $\langle \ \rangle_R$ by numerical integrations.

\begin{table}[Table I]
\begin{tabular}{rrrrr} 
$\delta\epsilon/\Delta\epsilon$\ \  &\ \ 3.2\ \ &\ \ 3.6\ \ &\ \
4.0\ \ &\ \ 4.4 \ \
\\ \hline
$\Delta\epsilon/T_0$\ \  & 0.370 & 0.336 & 0.309 & 0.286 \\
$\Delta\epsilon/T_f$\ \  & 0.339 &  0.304 & 0.276 & 0.252\\
$T_f/T_0$\ \  & 1.091 & 1.105 & 1.120 & 1.135 \\
\end{tabular}
\caption{Comparison between the thermodynamic folding
transition temperature
$T_f$ and the dynamic transition temperature $T_0$.}
\end{table}

\end{document}